\documentclass[11pt]{article}
\usepackage[pdftex]{graphicx,color} 
\usepackage{jheppub}
\usepackage{amsmath}
\usepackage{amssymb}
\usepackage{comment}
\usepackage{multirow}
\usepackage{mathtools}
\usepackage{dsdshorthand}
\usepackage{tikz}
\usepackage{shuffle}
\usetikzlibrary{arrows}
\tikzset{
>=stealth'
}
\newcommand{\bea}{\begin{equation}\begin{aligned}}
\newcommand{\eea}[1]{\label{#1}\end{aligned}\end{equation}}
\newcommand{\beq}{\begin{equation}}
\newcommand{\eeq}{\end{equation}}

\title{The AdS Virasoro-Shapiro Amplitude}

\author{Luis F. Alday and Tobias Hansen}

\affiliation{Mathematical Institute, University of Oxford,
Woodstock Road, Oxford, OX2 6GG, UK}

\abstract{
We present a constructive method to compute the AdS Virasoro-Shapiro amplitude, order by order in AdS curvature corrections. At $k^{\text{th}}$ order the answer takes the form of a genus zero world-sheet integral involving weight $3k$ single-valued multiple polylogarithms. The coefficients in our ansatz are fixed, order by order, by requiring: crossing symmetry; the correct supergravity limit; the correct structure of poles, determined by dispersive sum rules; and the dimensions of the first few Konishi-like operators, available from integrability. We explicitly construct the first two curvature corrections. Our final answer then reproduces all localisation results and all CFT data available from integrability, to this order, and produces a wealth of new CFT data for planar $\mathcal{N}=4$ SYM at strong coupling.
}

\emailAdd{alday@maths.ox.ac.uk, tobias.hansen@maths.ox.ac.uk}

\begin{document}

\maketitle

\section{Introduction}

Over the last few years our understanding and ability to compute super-string scattering amplitudes in flat space has significantly advanced. One of the earliest results in string theory is the tree-level amplitude for the scattering of four massless supergravity states (gravitons)
\begin{equation}
A_4(\varepsilon_i,p_i) = K(\varepsilon_i,p_i) \int dz^2 |z|^{-2S-2}|1-z|^{-2T-2}\,,
\end{equation}
where $S+T+U=0$ denote the Mandelstam variables and a simple prefactor $K(\varepsilon_i,p_i)$ encodes the dependence on the polarisation vectors of the gravitons. By now compact expressions exist for the tree-level scattering of gravitons with arbitrary multiplicity \cite{Mafra:2011nv,Stieberger:2013wea}. These amplitudes display remarkable structures and much has been learnt from them. For instance, tree-level open and closed-string amplitudes are related by the KLT relations \cite{Kawai:1985xq}, and this has inspired powerful relations between supergravity and Yang-Mills amplitudes \cite{Bern:2008qj}. From a mathematical viewpoint these relations can be understood as a single-valued map between tree-level open and closed superstring amplitudes \cite{Stieberger:2013wea,Brown:2019wna,Schlotterer:2018zce,Vanhove:2018elu}. More precisely, the low-energy expansion of closed-string amplitudes contains only single-valued multiple zeta values, a subclass of zeta values  \cite{Brown:2013gia} obtained by evaluating single-valued multiple polylogarithms \cite{Brown:2004ugm} at unity. Defining then a single-valued map 
\begin{equation}
\text{sv}: \zeta(n_1,n_2,\ldots) \to \zeta^{\text{sv}}(n_1,n_2,\ldots),
\end{equation}
it turns out this map takes open-string amplitudes to closed-string amplitudes, making manifest a surprising interplay with number theory. 

In contrast, string amplitudes in curved backgrounds are largely unexplored. Perturbatively they organise by a genus expansion, but even at tree-level a direct world-sheet approach is at the moment out of reach. While progress has been made for cases with pure background NS-NS B-field, see \cite{Maldacena:2001km}, the presence of Ramond-Ramond fields prevents the use of the RNS formulation. Alternative approaches are either very hard to quantise in curved backgrounds (Green-Schwarz formalism) or are not yet at the point where the computation of tree-level amplitudes is possible (pure-spinor formalism). In this paper we study the tree-level amplitude for four massless supergravity states in type IIB theory on $AdS_5 \times S^5$. While we don't know how to quantise perturbative string theory on this background, we will combine world-sheet intuition with several powerful tools available in this case. On one hand, the AdS/CFT duality relates this observable to a four-point correlator in ${\cal N}=4$ SYM at large central charge, which can be studied by CFT methods. On the other hand, it has been recently observed that single-valuedness of the low energy expansions is also a powerful guiding principle when perturbing around flat-space \cite{Alday:2022uxp,Alday:2022xwz,Alday:2023jdk}. These observations lead to a precise proposal for the structure of the tree-level amplitude on $AdS_5 \times S^5$ in a large-radius expansion around flat space, to all orders. 

To understand the structure of our proposal let us ignore super-symmetry for a moment and consider a four-point string amplitude in the Polyakov formulation
\begin{equation}
A_4(p_i) \sim \int {\cal D} X {\cal D} g \, e^{-S_P} V(p_1) V(p_2) V(p_3)  V(p_4)\,,
\label{path_integral}
\end{equation}
with the Polyakov action given by
\beq
S_P=\frac{1}{4 \pi \alpha'} \int d^2 \sigma
\sqrt{g} g^{\alpha \beta} \partial_\alpha X^\mu
\partial_\beta X^\nu G_{\mu \nu}(X)\,.
\label{S}
\eeq
Assuming $G_{\mu \nu}(X)$ is the metric of $AdS_5 \times S^5$ with radius $R$, and expanding around flat space, we obtain
\beq
G_{\mu \nu}(X) = \eta_{\mu \nu} + \frac{h_{\mu \nu}}{R^2}
+ \cdots \,, \qquad
h_{\mu \nu} \sim X_\mu X_\nu \sim \lim_{q \to 0} \frac{\partial^2}{\partial q^\mu \partial q^\nu} e^{i q \cdot X}\,.
\eeq
Plugging this expansion back into the path integral \eqref{path_integral}, we see that AdS curvature corrections have the same effect as the addition of vertex operators for soft gravitons, whose momenta are taken to zero after taking two derivatives. The flat space $4+k$ graviton amplitude, where $k$ of the gravitons are soft, can be analysed using the soft graviton theorem\,\cite{Sen:2017xjn}
\beq
A_{n+1} (p_1,\ldots,p_n,\e q) = \sum\limits_{i=1}^n
\left(\frac{1}{\e} \frac{\varepsilon_{\mu \nu} p_i^\mu p_i^\nu}{p_i \cdot q}
+ \frac{\varepsilon \cdot p_i \varepsilon_\mu q_\nu J^{\mu \nu}_i}{p_i \cdot q}
+O(\e)\right) A_{n} (p_1,\ldots,p_n)\,,
\eeq
where $J^{\mu \nu}_i$ is the angular momentum operator for particle $i$ and in particular a first-order differential operator on the momenta $p_i$. Schematically we then expect 
\begin{equation}
A_{4+k} (p_1,\ldots,p_4, \e q_1,\ldots,\e q_k) \sim \frac{1}{\e^k} A_{4} (p_1,\ldots,p_4) + \frac{1}{\e^{k-1}} {\cal D}_{1}A_{4} (p_1,\ldots,p_4) + \cdots\,,
\end{equation}
where $ {\cal D}_{1}$ is a first-order differential operator on the momenta $p_i$, $i=1,2,3,4$, whose precise form depends on the order in which the soft limits are taken. At the level of the integrand and recalling $A_4(p_i) \sim \int dz^2 |z|^{-2S-2}|1-z|^{-2T-2}$ we see that the leading term has the same form, while the subleading term is also of the same form, with the extra insertion of factors like $\log |z|^2$ and $\log |1-z|^2$: namely, single-valued polylogarithms of weight one. According to the discussion above, we further need to take $2k$ derivatives (two per extra soft graviton), which leads us to focus on the term proportional to $\e^{2k}$. This is expected to be of the form  
\begin{equation}
\int dz^2 |z|^{-2S-2}|1-z|^{-2T-2} {\cal L}_{|w|=3k}(z),
\end{equation}
namely, the usual genus-zero integral for the scattering of four gravitons, with the additional insertion of weight $3k$ single-valued multiple polylogarithms. This will be the basic building block for our construction. A few comments are in order. First, in the full computation divergent terms, as we take the momenta of the soft gravitons to zero, should cancel out. The model discussed here is simply a toy model. Second, on general grounds one would also expect terms of lower transcendentality at a given order, however, all solutions we have found have uniform and maximal transcendentality. This mimics what happens in other contexts in ${\cal N}=4$ SYM. Finally, note that the insertion of single-valued multiple polylogarithms will automatically produce single-valued zeta values  in the low energy expansion, upon integration over the Riemann sphere. 

The remainder of the paper is organised as follows.
In section \ref{sec:overview} we present our strategy for fixing the AdS Virasoro-Shapiro amplitude order by order in curvature corrections, which is then carried out in the subsequent sections.
Section \ref{sec:dispersive_sum_rules} contains the dispersive sum rules up to order $1/\lambda$, and section \ref{sec:world-sheet_correlator} discusses the ansatz for the world-sheet correlator and the solutions for the first two curvature corrections.
Having fixed these corrections gives us access to a wealth of new OPE data and Wilson coefficients, which are presented and where possible compared to previous results in section \ref{sec:data}.
Appendix \ref{app:sum_rules} contains additional definitions for the dispersive sum rules and appendix \ref{app:SVMPLs} discusses properties of single-valued multiple polylogarithms.

\section{Overview}
\label{sec:overview}
The central object of this paper is the tree-level amplitude of four gravitons in type IIB superstring theory on $AdS_5 \times S^5$, denoted the AdS Virasoro-Shapiro amplitude for short. By the AdS/CFT correspondence it is also the following correlator in $\mathcal{N}=4$ SYM, at leading non-trivial order in the large central charge expansion
\beq
\langle \mathcal{O}_2^{I_1 J_1} (x_1) \mathcal{O}_2^{I_2 J_2} (x_2) \mathcal{O}_2^{I_3 J_3} (x_3) \mathcal{O}_2^{I_4 J_4} (x_4) \rangle\,.
\label{correlator}
\eeq
Here $\mathcal{O}_2^{IJ}$ is the superconformal primary operator of the stress-tensor multiplet. It is a scalar operator with conformal dimension $\Delta=2$, transforming in the ${\bf 20'}$ representation of the $SU(4)$ $R-$symmetry group. The R-symmetry dependence of the correlator \eqref{correlator} is fully fixed by the superconformal Ward identities \cite{Dolan:2001tt}, and we can write \eqref{correlator} in terms of the {\it reduced correlator} $\mathcal{T}(U,V)$, which is a function of the conformal cross-ratios $U=\frac{x_{12}^2x_{34}^2}{x_{13}^2x_{24}^2}$, $V=\frac{x_{14}^2x_{23}^2}{x_{13}^2x_{24}^2}$. 

We will make use of the Mellin transform of the reduced correlator
\beq
  \cT(U, V)
   = \int_{-i \infty}^{i \infty} \frac{ds_1  ds_2}{(4 \pi i)^2} U^{\frac{s_1}{2}+\frac23} V^{\frac{s_2}{2} - \frac43}
    \Gamma \bigg(\frac43 - \frac{s_1}{2} \bigg)^2 \Gamma \bigg(\frac43 - \frac{s_2}{2} \bigg)^2 \Gamma \bigg(\frac43 - \frac{s_3}{2} \bigg)^2
    M(s_1, s_2) \,,
\label{mellin}
\eeq
where $s_1+s_2+s_3=0$. The low-energy expansion of the correlator takes a particularly simple form in Mellin space, containing the tree-level Witten diagrams of supergravity plus an infinite tower of contact diagrams with higher derivative quartic couplings
\begin{equation}
M(s_1,s_2) = \frac{8}{(s_1-\frac23)(s_2-\frac23)(s_3-\frac23)} + \sum\limits_{a,b=0}^\infty  \frac{\Gamma(2a+3b+6)}{8^{a+b}\lambda^{\frac32 + a + \frac32 b} } \sigma_2^a \sigma_3^b
\left(  \alpha^{(0)}_{a,b} + \frac{\alpha^{(1)}_{a,b} }{\sqrt{\lambda}}  + \frac{ \alpha^{(2)}_{a,b}}{\lambda}
 + \cdots \right)
\label{M}
\end{equation}
where $\sigma_2=s_1^2+s_2^2+s_3^2$ and $\sigma_3=s_1 s_2 s_3$.
This is an expansion in large t'Hooft coupling $\lambda$ where we keep all orders, which can equivalently be written in terms of $R$ and $\alpha'$ via the dictionary
\beq
\frac{1}{\sqrt{\lambda}} = \frac{\alpha'}{R^2}\,.
\eeq
As we will see below, the Wilson coefficients  $\alpha^{(0)}_{a,b}$ reproduce the flat space result (the usual Virasoro-Shapiro amplitude) while subsequent terms $\alpha^{(1)}_{a,b},\alpha^{(2)}_{a,b},\cdots$ give the $AdS$-curvature corrections. 

The reduced correlator admits a decomposition in terms of exchanged super-conformal primaries, and the Mellin amplitude generically has poles corresponding to these exchanges. The exchanged operators include both single and double-trace operators. While the poles of double-trace operators are already taken into account by the measure in \eqref{mellin}, one expects poles corresponding to the exchange of (heavy) single-trace operators. Furthermore, as a string amplitude, the Mellin amplitude also enjoys soft UV behaviour, namely a polynomial bound in the Regge limit, the bound on chaos \cite{Maldacena:2015waa}.
Together these facts were used in \cite{Alday:2022uxp,Alday:2022xwz} to derive dispersive sum rules that relate the Wilson coefficients $\alpha^{(k)}_{a,b}$ in \eqref{M} to the OPE data of the single-trace superconformal primary operators exchanged.

To make further contact with the string world-sheet we have to understand how to sum the low-energy expansion. While the sum in  \eqref{M} has zero radius of convergence, it turns out that it is Borel summable. For this reason we study the Borel transform of the Mellin amplitude
\beq \label{flat}
A(S,T) = 2 \lambda^\frac{3}{2} \int_{\kappa-i\infty}^{\kappa+ i \infty} \frac{d\alpha}{2 \pi i} \, e^\alpha \alpha^{-6} 
M \left( \frac{2\sqrt{\lambda} S}{\alpha}, \frac{2\sqrt{\lambda} T}{\alpha} \right)  \,.
\eeq
Note that at leading order this coincides with the flat space limit as introduced in \cite{Penedones:2010ue}. The Borel transform of the low-energy expansion \eqref{M} reads\footnote{Perturbatively in a $1/\lambda$ expansion, which is the regime we are working on in this paper. }
\bea
A(S,T) ={}& A^{(0)}(S,T) + \frac{1}{\sqrt{\lambda}} A^{(1)}(S,T)  + \frac{1}{\lambda} A^{(2)}(S,T) +\cdots \,,\\
A^{(k)}(S,T) ={}& \text{SUGRA}^{(k)} +
2\sum\limits_{a,b=0}^\infty  \hat\sigma_2^a \hat\sigma_3^b   \alpha^{(k)}_{a,b} \,,\\
\text{SUGRA}^{(0)} ={}& \frac{1}{\hat\sigma_3} \,, \quad
\text{SUGRA}^{(1)} = -\frac{2}{3} \frac{\hat\sigma_2}{\hat\sigma_3^2} \,, \quad
\text{SUGRA}^{(2)} = \frac{2}{9} \frac{\hat\sigma_2^2}{\hat\sigma_3^3} \,, \quad
\text{SUGRA}^{(k>2)} = 0\,,
\eea{A_expansion}
with $S+T+U=0$, $\hat\sigma_2= \frac12 ( S^2+T^2+U^2)$ and $\hat\sigma_3 = STU$.
The leading contribution is the Virasoro-Shapiro amplitude for type IIB superstring theory in flat space
\beq
A^{(0)}(S,T) =-\frac{ \Gamma \left(- S\right) \Gamma \left(-T\right) \Gamma \left(- U \right) }{\Gamma \left(S +1\right) \Gamma \left( T+1\right) \Gamma \left( U +1\right) }\,,
\label{VS}
\eeq
where an overall factor containing the graviton polarisations is stripped off because we are studying the reduced correlator.  In this interpretation we can identify the variables $S, T, U$ with the Mandelstam variables
\beq
S = - \frac{\a'}{4} (p_1+p_2)^2\,, \qquad
T = - \frac{\a'}{4} (p_1+p_3)^2\,, \qquad
U = - \frac{\a'}{4} (p_1+p_4)^2\,.
\eeq
The first AdS correction $A^{(1)}(S,T)$ was fully determined in \cite{Alday:2022uxp,Alday:2022xwz,Alday:2023jdk} and in the present paper we will describe an algorithm that allows us to determine further corrections. We will demonstrate this by fully fixing the next correction $A^{(2)}(S,T)$. To this end we make the assumption that $A^{(k)}(S,T)$ should also have a representation as an integral over the Riemann sphere, the world-sheet for genus 0 closed string amplitudes
\beq
A^{(k)}(S,T) = \int d^2 z |z|^{-2S-2}|1-z|^{-2T-2} G^{(k)}_{\text{tot}}(S,T,z)\,,
\label{worldsheet}
\eeq
where the integration measure is defined as $d^2 z =dz d \bar z /(-2 \pi i)$.
The flat space amplitude \eqref{VS} has this form with the manifestly crossing-symmetric integrand
\beq
G^{(0)}_{\text{tot}}(S,T,z) = \frac13 \left( \frac{1}{U^2} + \frac{|z|^2}{S^2} + \frac{|1-z|^2}{T^2} \right) \,.
\label{G0}
\eeq
At general order we expect the following structure
\beq
G^{(k)}_{\text{tot}}(S,T,z) = G^{(k)}\left(S,T,z\right) + |z|^2 G^{(k)}\left(U,T,\tfrac{1}{z}\right)+ |1-z|^2 G^{(k)}\left(S,U,\tfrac{z}{z-1}\right),
\eeq
with $G^{(k)}(S,T,z)$ symmetric under the simultaneous exchange $z \to 1-z$ and $S \leftrightarrow T$, so that after integration a symmetric function in $S,T$ is produced. In our proposed solution $G^{(k)}(S,T,z)$ is a single-valued function of $z$ of transcendental weight $3k$, as motivated in the introduction, and rational in $S,T$ with homogeneous degree $2k-2$. More precisely
\beq
G^{(k)}(S,T,z) = \sum_u \frac{p^{(k)s}_u(S,T)}{U^2}{{\cal L}_u^{(k)s}}(z)+ \sum_v \frac{p^{(k)a}_v(S,T)}{U^2}{{\cal L}_v^{(k)a}}(z)\,,
\eeq
where $u/v$ run over a basis of transcendentality $3k$ single-valued multiple polylogarithms (SVMPLs), symmetric/anti-symmetric under the exchange of $z \leftrightarrow 1-z$. This includes SVMPLs of  weight $3k$, but also $\zeta(3)$ times SVMPLs of  weight $3k-3$ and so on. Furthermore, $p^{(k)s}_u(S,T)$/$p^{(k)a}_v(S,T)$ are symmetric/anti-symmetric polynomials of degree $2k$. A solution for $G^{(1)}(S,T,z)$ was presented in \cite{Alday:2023jdk}.

At each order $k$, $G^{(k)}(S,T,z)$ depends on a finite number of coefficients. Our algorithm to fix them is to plug our ansatz in  \eqref{worldsheet} and compute the residues of $A^{(k)}(S,T)$
at $S=\delta$:
\beq
A^{(k)}(S,T) = \frac{R^{(k)}_{3k+1}(T,\delta)}{(S-\delta)^{3k+1}} + \frac{R^{(k)}_{3k}(T,\delta)}{(S-\delta)^{3k}} + \ldots + \frac{R^{(k)}_{1}(T,\delta)}{S-\delta} + \text{regular}\,, \quad \delta = 1,2,\ldots
\label{R_def}
\eeq
The same residues can be computed independently in terms of the OPE data of the exchanged single-trace operators, using the dispersive sum rules. At a given order, the higher order poles are fixed in terms of the OPE data at lower orders. This results in strong constraints for the coefficients in our ansatz, fixing $A^{(k)}(S,T)$ almost completely. For the cases we analysed, we found that $A^{(1)}(S,T)$ is actually fully fixed, while $A^{(2)}(S,T)$ is fully fixed once we input the conformal dimension of the Konishi operator at this order, available from integrability. In general we expect that providing the conformal dimensions of the first few Konishi-like operators, i.e.\ operators on the leading Regge trajectory, will fully fix $A^{(k)}(S,T)$. In the remaining sections we demonstrate our program explicitly for $A^{(1)}(S,T)$ and $A^{(2)}(S,T)$. This solution then passes various independent checks. In particular it matches the conformal dimensions of the whole tower of Konishi-like operators, obtained using integrability, and reproduces the two Wilson coefficients that were previously known from localisation.

\section{Dispersive sum rules}
\label{sec:dispersive_sum_rules}

The role of dispersive sum rules is to connect our expressions for $A^{(k)}(S,T)$ with the OPE data of single-trace superconformal primaries in the expansion
\beq
\mathcal{T}(U,V) = U^{-2} \sum\limits_{\substack{\text{superconformal}\\\text{primaries } \mathcal{O}_{\tau,\ell}}}
C^2_{\tau,\ell}G_{\tau+4,\ell}(U,V)\,,
\label{OPE}
\eeq 
where $G_{\tau,\ell}(U,V)$ is a conformal block in 4 dimensions which here takes into account the contributions for all (super-)descendants of a given superconformal primary\footnote{See \cite{Alday:2022uxp} for our definitions for the superconformal Ward identity and conformal block.}.
The leading conformal dimensions and OPE coefficients of these operators in a large $\lambda$ expansion are determined by the corresponding flat space data, so the operators can be labelled by the flat space mass level $\delta=1,2,\ldots$ as well as the spin $\ell$. In terms of these labels the operators are degenerate and the degeneracies were recently estimated in \cite{Alday:2023flc}. We show the expected degeneracies for the correlator at hand in Figure \ref{fig:degeneracies}.
\begin{figure}
\centering
  \begin{tikzpicture}[xscale=0.5,yscale=0.35]
    \coordinate (nw) at (-.5,12);
    \coordinate (sw) at (-.5,-1);
    \coordinate (se) at (14,-1);
    \draw[->] (sw) --  (nw) ;
    \draw[->] (sw) --  (se) ;
    \node at (0,-2) [] {$1$}; 
    \node at (2,-2) [] {$2$}; 
    \node at (4,-2) [] {$3$}; 
    \node at (6,-2) [] {$4$}; 
    \node at (8,-2) [] {$5$}; 
    \node at (10,-2) [] {$6$}; 
    \node at (12,-2) [] {$7$}; 
    \node at (14,-2) [] {$\delta$}; 
    \node at (-1,0) [] {$0$};
    \node at (-1,2) [] {$2$};
    \node at (-1,4) [] {$4$};
    \node at (-1,6) [] {$6$};
    \node at (-1,8) [] {$8$};
    \node at (-1,10) [] {$10$};
    \node at (-1,12) [] {$\ell$};
    \node at (0,0) [] {$1$};
    \node at (2,0) [] {$2$};
    \node at (2,2) [] {$1$};
    \node at (4,0) [] {$6$};
    \node at (4,2) [] {$4$};
    \node at (4,4) [] {$1$};
    \node at (6,0) [] {$22$};
    \node at (6,2) [] {$24$};
    \node at (6,4) [] {$6$};
    \node at (6,6) [] {$1$};
    \node at (8,0) [] {$99$};
    \node at (8,2) [] {$157$};
    \node at (8,4) [] {$40$};
    \node at (8,6) [] {$6$};
    \node at (8,8) [] {$1$};
    \node at (10,0) [] {$547$};
    \node at (10,2) [] {$1104$};
    \node at (10,4) [] {$331$};
    \node at (10,6) [] {$52$};
    \node at (10,8) [] {$6$};
    \node at (10,10) [] {$1$};
    \node at (12,0) [] {$3112$};
    \node at (12,2) [] {$7365$};
    \node at (12,4) [] {$2570$};
    \node at (12,6) [] {$461$};
    \node at (12,8) [] {$58$};
    \node at (12,10) [] {$6$};
    \node at (12,12) [] {$1$};
  \end{tikzpicture}
\caption{Degeneracies of superconformal primaries with $\tau_0=2\sqrt{\delta}$, even spin $\ell$ and $R$ charge 0 \cite{Alday:2023flc}.}
\label{fig:degeneracies}
\end{figure}
We define the large $\lambda$ expansion of the OPE data as
\begin{align}
\tau &= \tau_0 \lambda^{\frac{1}{4}} + \tau_1  +  \tau_2 \lambda^{-\frac{1}{4}} + \ldots \,, \label{twistsStringy}\\
C^2_{\tau,\ell} &=  \frac{\pi ^3}{2^{12}} \frac{ 2^{-2 \tau} \tau^6 }{\sin^2(\frac{\pi \tau}{2}) } \frac{1}{2^{2 \ell}(\ell+1)}  
\left( f_0 +  f_1 \lambda^{-\frac{1}{4}} + f_2 \lambda^{-\frac{1}{2}} + \ldots \right)\,. \label{OPEStringy1}
\end{align}
The dispersive sum rules for the correlator at hand up to order $1/\sqrt{\lambda}$ were derived in \cite{Alday:2022uxp,Alday:2022xwz} and we repeat them here for completeness
\begin{align}
\alpha^{(0)}_{a,b} ={}& 
\sum\limits_{\delta=1}^\infty \sum\limits_{q=0}^b
\frac{c_{a,b,q}}{\delta^{3+2a+3b}} F^{(0)}_{q}(\delta)\,,\nonumber\\
\alpha^{(1/2)}_{a,b} ={}& 
\sum\limits_{\delta=1}^\infty \sum\limits_{q=0}^b
\frac{c_{a,b,q}}{\delta^{\frac72+2a+3b}} \left(F^{(1)}_{q}(\delta) - (3+2a+3b) T^{(1)}_{q}(\delta)\right)\,,\label{alpha1}\\
\alpha^{(1)}_{a,b} ={}& 
\sum\limits_{\delta=1}^\infty \sum\limits_{q=0}^b
\frac{c_{a,b,q}}{\delta^{4+2a+3b}} \bigg(F^{(2)}_{q}(\delta) - (3+2a+3b) T^{(2)}_{q}(\delta) + \sum_{j=0}^1 (q+1)_j P^{(1)}_{3,j}(a,b,q) F^{(0)}_{q+j}(\delta)\bigg),\nonumber
\end{align}
Here $c_{a,b,q}$ are the combinatorical coefficients
\bea
c_{a,b,q} ={}& \frac{(-1)^q (2 a+3 b-3 q) \Gamma (a+b-q) }{2 \Gamma (a+1) \Gamma
   (b-q+1)}\\
& {}_4F_3\left(\tfrac{q+1}{2},\tfrac{q}{2},q-b,q+1-\tfrac{2
   }{3}a-b;q+1,q+1-a-b,q-\tfrac{2}{3}a-b;4\right)\,,
\eea{c}
and the functions $F^{(k)}_q(\delta)$ and $T^{(k)}_q(\delta)$ encode the OPE data, starting with
\beq
F^{(0)}_q (\delta) = \frac{4^q}{\Gamma(2q+2)} \sum_{\ell=0,2,\ldots}^{2(\de-1)} (\ell-q+1)_q (\ell+2)_q \langle f_0 \rangle_{\delta,\ell}\,,
\eeq
for the leading contribution to the OPE coefficients \eqref{OPEStringy1}.
The angle brackets $\langle \ldots \rangle_{\delta,\ell}$ denote a sum over all the degenerate OPE data for a given $\delta, \ell$, for which the degeneracies are shown in Figure \ref{fig:degeneracies}.
The precise definitions of the remaining functions  $F^{(k)}_q(\delta)$ and $T^{(k)}_q(\delta)$ along with the $P^{(k)}_{i,j}(a,b,q)$, which are polynomials in $a,b,q$ of degree $i-j$, are given in appendix \ref{app:sum_rules}.

The coefficients $\alpha^{(0)}_{a,b}$ are known from flat space \eqref{VS} and can be used to compute the OPE data $\tau_0(\delta,\ell) = 2 \sqrt{\delta}$ and $\langle f_0 \rangle_{\delta,\ell}$. The first corrections are somewhat trivial as matching with \eqref{M} yields
\beq
\alpha^{(1/2)}_{a,b} = 0\,, \quad \forall a,b\,,
\eeq
which has the solution
\beq
\tau_1(\delta,\ell) = -\ell-2\,, \qquad
\< f_1 \>_{\delta,\ell} =   \frac{3 \ell + \frac{23}{4}}{\sqrt{\delta}} \<f_0\>_{\delta,\ell}\,.
\eeq
These equations are used to eliminate $\tau_1(\delta,\ell)$ and $\< f_1\>_{\delta,\ell}$ from all the other dispersive sum rules. The sum rule for $\alpha^{(1)}_{a,b}$ was solved in \cite{Alday:2022xwz} and the solution fixes the OPE data $\langle f_0 \tau_2 \rangle_{\delta,\ell}$ and $\langle f_2 \rangle_{\delta,\ell}$.

The most efficient way to obtain the general sum rules is to use a crossing-symmetric dispersion relation as described in \cite{Alday:2022xwz}.
Performing this computation up to order $1/\lambda$ we find the following two new sum rules
\bea
{}&\alpha^{(3/2)}_{a,b} = \sum\limits_{\delta=1}^\infty 
  \sum\limits_{q=0}^b
\frac{c_{a,b,q}}{\delta^{\frac92+2a+3b}} \bigg( 
  F^{(3)}_q(\delta) - (3+2a+3b)T^{(3)}_q(\delta)
+ \sum_{j=0}^1 (q+1)_j P^{(3/2)}_{2,j}(q) F^{(0)}_{q+j}(\delta) \bigg)\,,\\
{}&\alpha^{(2)}_{a,b} = \sum\limits_{\delta=1}^\infty 
  \sum\limits_{q=0}^b
\frac{c_{a,b,q}}{\delta^{5+2a+3b}} \bigg( 
  F^{(4)}_q(\delta) - (3+2a+3b)T^{(4)}_q(\delta)
+ \frac{1}{4} (3+2 a+3 b) (7+4 a+6 b) T^{(2,2)}_q(\delta) \\
&+ \sum_{j=0}^1 (q+1)_j \left(P^{(2)}_{3,j}(a,b,q) F^{(2)}_{q+j}(\delta)
+ P^{(2)}_{4,j}(a,b,q) T^{(2)}_{q+j}(\delta)\right)
+ \sum_{j=0}^2 (q+1)_j P^{(2)}_{6,j}(a,b,q) F^{(0)}_{q+j}(\delta)
 \bigg)\,,
\eea{alpha2}
where further definitions can again be found in appendix \ref{app:sum_rules}.
We solve
\beq
\alpha^{(3/2)}_{a,b} = 0\,, \quad \forall a,b\,,
\eeq
to find
\beq
\tau_3(\delta,\ell) = 0\,, \quad
\langle f_3 \rangle_{\delta,\ell} = 
\frac{3 \ell + \frac{23}{4}}{\sqrt{\delta}} \left( \<f_2\>_{\delta,\ell} - \frac{\<f_0 \tau_2\>_{\delta,\ell}}{2 \sqrt{\delta}} \right)
-\frac{560 \ell^3+3220 \ell^2+6164 \ell+3931}{64 \delta ^{3/2}}\<f_0\>_{\delta,\ell}
\,.
\eeq
Finally, the sum rule for $\alpha^{(2)}_{a,b}$ contains unknown data on both sides of the equation, in particular the OPE data
\beq
F^{(4)}_q(\delta) \leftrightarrow \langle f_4 \rangle_{\delta,\ell}\,, \qquad
T^{(4)}_q(\delta) \leftrightarrow\langle f_0 \tau_4 + f_2 \tau_2 \rangle_{\delta,\ell}\,, \qquad
T^{(2,2)}_q(\delta) \leftrightarrow\langle f_0 \tau_2^2 \rangle_{\delta,\ell}\,.
\label{new_OPE_data}
\eeq
In order to relate the integral representation for $A^{(2)}(S,T)$ to this OPE data we would like to resum the low-energy expansion
\beq
2 \sum_{a,b=0}^{\infty} \hat\sigma_2^a \hat\sigma_3^b   \alpha^{(2)}_{a,b}
= \sum\limits_{\delta=1}^\infty 
  \sum\limits_{q=0}^{2(\delta-1)}
\sum_{a,b=0}^{\infty} P_{\delta,q}(a,b) c_{a,b,q} \left(\frac{\hat\sigma_2}{\delta^2}\right)^a \left( \frac{\hat\sigma_3}{\delta^3}\right)^b\,,
\eeq
where $P_{\delta,q}(a,b)$ are polynomials in $a$ and $b$ and the sum over $q$ truncates  for fixed $\delta$ because
\beq
T^{(k)}_q(\delta) = 0\,, \qquad F^{(k)}_q(\delta) = 0 \,, \qquad q > 2(\delta-1)\,.
\eeq
The sums over $a$ and $b$ can now be done by using \cite{Zagier:2019eus}
\beq
\sum\limits_{a,b=0}^\infty c_{a,b,q} x^a y^b =
\frac12 \frac{y+2}{1-x-y} \left(\frac{\sqrt{1-4y}-1}{2} \right)^q\,,
\eeq
and turning  $a$ and $b$ into operators $x \partial_x$ and  $y \partial_y$ acting on this sum.
In this way we can compute the residues at $S=\delta$ defined in \eqref{R_def}.
The residues of order seven to four are completely determined in terms of known OPE data, whereas the residues of order three to one depend on the OPE data \eqref{new_OPE_data}. The expressions for all the residues of $A^{(2)}(S,T)$ can be found in \eqref{R_from_dsr_7-4} and \eqref{R_from_dsr_3-1}.
The fact that the dispersive sum rules fix most of $A^{(k)}(S,T)$ in terms of OPE data from lower orders is one of the reasons we are able to solve for $A^{(k)}(S,T)$ order by order.

\section{World-sheet correlator}
\label{sec:world-sheet_correlator}

We start with the following ansatz for the world-sheet correlator
\begin{equation}
A^{(k)}(S,T)= B^{(k)}(S,T)+ B^{(k)}(U,T)+B^{(k)}(S,U)\,,
\end{equation}
where $B^{(k)}(S,T)$ is symmetric under exchange of $S$ and $T$ and has the representation
\begin{equation}
B^{(k)}(S,T)=\int d^2 z |z|^{-2S-2}|1-z|^{-2T-2} G^{(k)}(S,T,z)\,,
\label{B}
\end{equation}
so that $A^{(k)}(S,T)$ is manifestly crossing-symmetric. The symmetry condition for $B^{(k)}(S,T)$ translates to the following condition for $G^{(k)}(S,T,z)$
\beq
G^{(k)}(S,T,z) =  G^{(k)}(T,S,1-z)\,.
\label{G_ST_sym}
\eeq
We further impose $G^{(k)}(S,T,z)$ to be even under the exchange $z \leftrightarrow \bar{z}$ because it should be a function of the 2D conformal cross-ratios $z \bar{z}$ and $(1-z)(1-\bar{z})$
\beq
G^{(k)}(S,T,z) =  G^{(k)}(S,T,\bar{z})\,.
\label{G_zzb_sym}
\eeq
Note that the integral \eqref{B} would project out the odd component anyway.

\subsection{Basis of SVMPLs}

Let us construct a basis of SVMPLs that will allow us to easily solve \eqref{G_ST_sym} and \eqref{G_zzb_sym}.
For words of length $L$ there are $2^L$ linearly independent SVMPLs. To solve \eqref{G_zzb_sym} we would like to project on the even component under the exchange $z \leftrightarrow \bar z$. Given a word $w$ we denote by $\tilde w$ the reverse word. The projection is then equivalent to a quotient by the equivalence 
\begin{equation}
{\cal L}_{w}(z) \simeq {\cal L}_{\tilde w}(z) \,.
\end{equation}
More precisely, it can be shown that the even components of ${\cal L}_{w}(z) $ and ${\cal L}_{\tilde w}(z)$ are the same, up to single-valued multiple zeta values times SVMPLs of lower weight. At length $L$ this projection reduces the number of independent SVMPLs from $2^L$ to $2^{L-1}+2^{\left\lfloor \frac{L-1}{2}\right\rfloor}$. 
For instance at length three we get from $8 \to 6$ while at length six we get from $64 \to 36$. In our ansatz at weight three we also need to include $\zeta(3)$ (times the polylogarithm 1), giving a total of 7 independent functions. At weight six we also include $\zeta(3)$ times the projection for length three (6 functions), plus $\zeta(5)$ times the projection for length one (2 functions) plus $\zeta(3)^2$ times $1$ to get in total $36+6+2+1=45$ independent functions. 

SVMPLs are also closed under the transformation $z \to 1-z$ and we find it convenient to split them into symmetric and anti-symmetric components, defining
\bea
\mathcal{L}^s_{w}(z) ={}& \mathcal{L}_{w}(z) + \mathcal{L}_{w}(1-z)+\mathcal{L}_{w}(\bar{z}) + \mathcal{L}_{w}(1-\bar{z})\,,\\
\mathcal{L}^a_{w}(z) ={}& \mathcal{L}_{w}(z) - \mathcal{L}_{w}(1-z)+\mathcal{L}_{w}(\bar{z}) - \mathcal{L}_{w}(1-\bar{z})\,.
\eea{L_symmetrised}
At the level of the words (and modulo zeta values times SVMPLs of lower order), the transformation $z \to 1-z$ flips $0 \leftrightarrow 1$. For the total ansatz at transcendentality three we have 4 symmetric and 3 anti-symmetric functions
\bea
\mathcal{L}^{(1)s} ={}& \Big(
\mathcal{L}^s_{000}(z),
\mathcal{L}^s_{001}(z),
\mathcal{L}^s_{010}(z),
\zeta(3)
\Big)\,,\\
\mathcal{L}^{(1)a} ={}& \Big(
\mathcal{L}^a_{000}(z),
\mathcal{L}^a_{001}(z),
\mathcal{L}^a_{010}(z)
\Big)\,,
\eea{k1_L_basis}
and at transcendentality six (for which we have 45 functions), we have 25 symmetric and 20 anti-symmetric functions
\begin{align}
\mathcal{L}^{(2)s} ={}& \Big(
\mathcal{L}^s_{000000}(z),
\mathcal{L}^s_{000001}(z),
\mathcal{L}^s_{000010}(z),
\mathcal{L}^s_{000011}(z),
\mathcal{L}^s_{000100}(z),
\mathcal{L}^s_{000101}(z),
\mathcal{L}^s_{000110}(z),\nonumber\\&
\mathcal{L}^s_{000111}(z),
\mathcal{L}^s_{001001}(z),
\mathcal{L}^s_{001010}(z),
\mathcal{L}^s_{001011}(z),
\mathcal{L}^s_{001100}(z),
\mathcal{L}^s_{001101}(z),
\mathcal{L}^s_{001110}(z),\nonumber\\&
\mathcal{L}^s_{010001}(z),
\mathcal{L}^s_{010010}(z),
\mathcal{L}^s_{010101}(z),
\mathcal{L}^s_{010110}(z),
\mathcal{L}^s_{011001}(z),
\mathcal{L}^s_{011110}(z),\nonumber\\&
\zeta(3)\mathcal{L}^s_{000}(z),
\zeta(3)\mathcal{L}^s_{001}(z),
\zeta(3)\mathcal{L}^s_{010}(z),
\zeta(5)\mathcal{L}^s_{0}(z),
\zeta(3)^2\Big)\,,
\label{L_s}
\end{align}
and
\begin{align}
\mathcal{L}^{(2)a} ={}& \Big(
\mathcal{L}^a_{000000}(z),
\mathcal{L}^a_{000001}(z),
\mathcal{L}^a_{000010}(z),
\mathcal{L}^a_{000011}(z),
\mathcal{L}^a_{000100}(z),
\mathcal{L}^a_{000101}(z),
\mathcal{L}^a_{000110}(z),\nonumber\\&
\mathcal{L}^a_{001001}(z),
\mathcal{L}^a_{001010}(z),
\mathcal{L}^a_{001100}(z),
\mathcal{L}^a_{001101}(z),
\mathcal{L}^a_{001110}(z),
\mathcal{L}^a_{010001}(z),
\mathcal{L}^a_{010010}(z),\nonumber\\&
\mathcal{L}^a_{010110}(z),
\mathcal{L}^a_{011110}(z),
\zeta(3)\mathcal{L}^a_{000}(z),
\zeta(3)\mathcal{L}^a_{001}(z),
\zeta(3)\mathcal{L}^a_{010}(z),
\zeta(5)\mathcal{L}^a_{0}(z)\Big)\,.
\label{L_a}
\end{align}
In terms of these vectors of basis elements our ansatz reads
\beq
G^{(k)}(S,T,z) = \sum\limits_{u} r^{(k)s}_{u} \mathcal{L}^{(k)s}_u +\sum\limits_{v} r^{(k)a}_{v} \mathcal{L}^{(k)a}_v \,,
\eeq
where $r^{(k)s}_{u} / r^{(k)a}_{v}$ are symmetric / antisymmetric homogeneous functions of $S$ and $T$ of weight $2k-2$.

\subsection{Ambiguities}

Since $S+T+U=0$, certain insertions $G^{(k)}(S,T,z)$ contribute to $B^{(k)}(S,T)$ but don't contribute to the total answer $A^{(k)}(S,T)$. This leads to an ambiguity when constructing integrands. Since we are ultimately interested in the final answer $A^{(k)}(S,T)$, integrands differing by these ambiguities are of course equivalent. Let us start with weight zero. At this order 
\begin{equation}
G^{(0)}(S,T,z) = f(S,T)\,,
\end{equation}
and symmetry in $S,T$ implies $f(S,T)=f(T,S)$. Any function $f(S,T)$ satisfying
\begin{equation}
U^2(f(S,T)+f(T,S)) + T^2(f(S,U)+f(U,S))+S^2(f(U,T)+f(T,U))=0\,,
\label{k0ambiguity}
\end{equation}
leads to a vanishing contribution to $A^{(0)}(S,T)$.
If we now assume that $f(S,T)$ has the correct denominator
\beq
f(S,T) = \frac{c}{U^2}\,,
\eeq
the constraint \eqref{k0ambiguity} becomes $c=0$ and there is no more ambiguity.
For higher weights we will also fix the denominator to be $1/U^2$, mimicking the leading term \eqref{G0}
\beq
r^{(k)s}_{u} = \frac{p^{(k)s}_{u}(S,T)}{U^2}\,, \qquad
r^{(k)a}_{v} = \frac{p^{(k)a}_{v}(S,T)}{U^2}\,.
\eeq
This fixes part of the ambiguities, but in general a finite number of ambiguities remains even after fixing the denominator.
Our final result for $G^{(k)}(S,T,z)$ then has the form
\beq
G^{(k)}(S,T,z) = \sum\limits_{u} r^{(k)s}_{u} \mathcal{L}^{(k)s}_u +\sum\limits_{v} r^{(k)a}_{v} \mathcal{L}^{(k)a}_v 
+ \sum_{j=1}^{n_{\text{amb}}} a_j \left( \sum\limits_{u} \hat{r}^{(k)s}_{ju} \mathcal{L}^{(k)s}_u +\sum\limits_{v} \hat{r}^{(k)a}_{jv} \mathcal{L}^{(k)a}_v  \right) \,,
\eeq
in terms of $n_{\text{amb}}$ unfixed coefficients $a_j$ which parameterise the ambiguities.
For the case $k=1$ there are two ambiguities, given by
\beq
\hat{r}^{(1)s}_1 = \left(0,0,0, \frac{S^2 + 4ST+ T^2}{(S+T)^2} \right)\,, \qquad
\hat{r}^{(1)a}_1 = \left(0,0,0\right)\,,
\eeq
and
\bea
\hat{r}^{(1)s}_2 &= \left(-\frac{5 S^2+8 S T+5 T^2}{12 (S+T)^2},\frac{S^2+S T+T^2}{3 (S+T)^2},-\frac{5 S^2+8 S T+5 T^2}{12 (S+T)^2},\frac{2S^2+2T^2}{(S+T)^2}\right)\,, \\
\hat{r}^{(1)a}_2 &= \frac{S-T}{S+T} \left(-\frac{5}{12},\frac{1}{3},\frac{1}{4}\right)\,.
\eea{A1_amb2}
For $k=2$ we
find a 17 parameter family of ambiguities, spanned by complicated vectors. The full set of ambiguities is included in the ancillary Mathematica notebook.  

\subsection{Solutions}

In order to fix the coefficients in our ansatz, we 
compute the residues of $A^{(k)}(S,T)$ using \eqref{worldsheet}, recalling that
\begin{equation}
G^{(k)}_{\text{tot}}(S,T,z) = G^{(k)}(S,T,z)+|z|^2 G^{(k)}(U,T,\tfrac{1}{z}) 
+|1-z|^2 G^{(k)}(S,U,\tfrac{z}{z-1})\,.
\end{equation}
Single-valued polylogarithms are closed under  $z \to 1-z$, $z \to 1/z$ and $z \to \frac{z}{z-1}$ and we explain how to compute their transformation properties in appendix \ref{app:SVMPLs}. This can be used to write $G^{(k)}_{\text{tot}}(S,T,z)$ in terms of SVMPLs of argument $z$, which are easy to expand around $z=0$. We can then compute any residue at $S=\delta=1,2,\ldots$ by integrating over a disc around $z=0$ using polar coordinates $z=\rho e^{i\alpha}$. The poles then arise through integrals of the form
\beq
\int_{0}^{\rho_0} d\rho \, \rho^{-2S+2\delta-1} \frac{1}{p!} \log^p (\rho^2)
= -\frac12 \frac{1}{(S-\delta)^{p+1}} + O\left( (S-\delta)^0 \right)\,.
\eeq
The residues at $S=0$ can be computed using the polar terms in the low energy expansion as discussed in section \ref{sec:wilson_coefficients} below. These poles are matched to the supergravity terms shown in \eqref{A_expansion}.

For $k=1$ our ansatz has $2 \cdot 4 + 3$ rational parameters, which are fixed by matching the residues with those from the SUGRA term and the dispersive sum rules,
where it is not necessary to specify the OPE data  $\langle f_0 \tau_2 \rangle_{\delta,\ell}$ or $\langle f_2 \rangle_{\delta,\ell}$. The result reads
\beq
r^{(1)s} = \left(- \frac16\,,  0\,, - \frac14\,,  2 \right)\,, \qquad
r^{(1)a} = \frac{S-T}{S+T} \left(- \frac16\,,  \frac13 \,,  \frac16  \right)\,.
\eeq

For the next correction at $k=2$ our ansatz for the functions $r^{(2)s}_u$ and $r^{(2)a}_v$ can be parametrised by $3 \cdot 25 + 2 \cdot 20 = 115$ rational numbers. 17 of them are ambiguities, so that $A^{(2)}(S,T)$ depends on 98 coefficients. We can now match the residues from the ansatz with those of the supergravity term and with the expressions \eqref{R_from_dsr_7-4} and \eqref{R_from_dsr_3-1} computed from the dispersive sum rules, making the assumption that the OPE data has the form
\bea
\langle f_4 \rangle_{\delta,\ell} ={}&  q^1_{\delta,\ell} \zeta(3)^2 + q^2_{\delta,\ell} \zeta(5) + q^3_{\delta,\ell} \zeta(3) + q^4_{\delta,\ell}  \,,& \\
\langle f_0 \tau_4 + f_2 \tau_2 \rangle_{\delta,\ell} ={}& q^5_{\delta,\ell} \zeta(3) + q^6_{\delta,\ell} \,, & \qquad q^i_{\delta,\ell} \in \mathbb{Q}\,,\\
\langle f_0 \tau_2^2 \rangle_{\delta,\ell}={}& q^7_{\delta,\ell}\,.&
\eea{OPE_zeta_assumption}
Matching all residues (of order 1 to 7) for $\delta=0,1,\ldots,6$ fixes 94 of the 98 parameters and including also the cases $\delta=7,8$ does not fix any more parameters. The remaining four parameters can be fixed using the
fact that the operators on the leading Regge trajectory are non-degenerate, as well as the known dimension $\tau_4(1,0)$ of the Konishi operator, to insert the OPE data
\beq
\langle f_0 \tau_2^2 \rangle_{1,0} = 4 \,, \qquad
\langle f_0 \tau_2^2 \rangle_{2,2} = \frac{27}{2}\,, \qquad
\langle f_0 \tau_4 + f_2 \tau_2 \rangle_{1,0} = \zeta(3) + \frac{413}{16}\,.
\eeq
The final result has the form
\begin{align}
r^{(2)s} ={}& \frac{S^2 + T^2}{2^4 3^5} \Big(-216,26739,13111,-7271,-9286,-9139,-26100,9219,-12672,-15917,\nonumber\\
&3541,-9901,-823,29697,-17307,1674,10530,3780,23760,-3483,0,0,0,0,0 \Big)\nonumber\\
& + \frac{S T}{2^4 3^5} \Big( 
216,8163,24433,-132845,-33460,-92347,-25725,67200,-21045,18571,\nonumber\\
&27967,-7363,52694,9372,1848,-7575,21006,26760,26769,55233,0,0,0,0,0
\Big)\,,\nonumber\\
r^{(2)a} ={}& \frac{(S^2 + T^2)(S-T)}{2^4 3^5 (S+T)} \Big(
-216,26739,13111,-7271,-9286,-9139,-26100,-12672,\nonumber\\
&-15917,-9901,2417,17061,-17307,1674,-432,3483,0,0,0,0
\Big)\label{r2}\\
&+ \frac{ST(S-T)}{2^4 3^5 (S+T)} \Big(
-216,61641,50655,-56985,-52032,4521,-26307,-2387,\nonumber\\
&-21173,5559,-43268,-16916,-67642,-19393,11432,15345,0,0,-84456,-5292
\Big)\,,\nonumber
\end{align}
where we used the ambiguities to cancel the $1/U^2$ in $r^{(2)s}$ and to set some of the entries to zero.

\section{Data and checks}
\label{sec:data}

\subsection{OPE data}
\label{sec:ope_data}

After having completely fixed $A^{(2)}(S,T)$ we can now compute the OPE data for any mass level $\delta$ by computing the residues of $A^{(2)}(S,T)$ at $S=\delta$ and matching them with the expressions from the sum rules \eqref{R_from_dsr_3-1}. We include all the OPE data for $\delta \leq 13$ in a Mathematica notebook.
Along the Regge trajectories the OPE data admits analytic formulas, which we obtain by matching to our data. For the first Regge trajectory we find
\begin{align}
\langle f_0 \tau_2^2 \rangle_{\delta,2(\delta-1)} ={}& \frac{r_0(\delta)}{4 \delta ^2} \left(3 \delta ^2-\delta +2\right)^2 \,, \nonumber\\
\langle f_0 \tau_4 + f_2 \tau_2  \rangle_{\delta,2(\delta-1)} ={}& r_0(\delta) \sqrt{\delta } \left(3 \delta ^2-\delta -1\right) \zeta (3)\label{OPE_data_n=0}\\
&-\frac{r_0(\delta)}{192 \delta ^{5/2}} \left(336 \delta ^5-5476 \delta ^4+2984 \delta ^3-3689 \delta ^2+439 \delta +450\right)\,, \nonumber\\
\langle f_4 \rangle_{\delta,2(\delta-1)} ={}&
r_0(\delta) \bigg( 2 \delta^3 \zeta(3)^2 - 3 \delta^2 \zeta(5) 
+ \frac{1}{48} \left(-112 \delta ^3+1728 \delta ^2-584 \delta -345\right) \zeta(3)\nonumber\\
&+\frac{49 \delta ^3}{72}-\frac{389 \delta ^2}{20}+\frac{4831 \delta }{72}-\frac{7411}{192}-\frac{16415}{288 \delta } +\frac{13219}{1920 \delta ^2}+\frac{6723}{2048 \delta ^3}\bigg)\,,\nonumber
\end{align}
with
\beq
r_n(\delta) = \frac{4^{2-2 \delta } \delta ^{2 \delta -2 n-1} (2 \delta -2 n-1)}{\Gamma (\delta ) \Gamma \left(\delta -\left\lfloor \frac{n}{2}\right\rfloor \right)}\,.
\eeq
As these operators are non-degenerate (see Figure \ref{fig:degeneracies}) we can omit the angle brackets and solve for the twists
\bea
\tau\left(\tfrac{\ell}{2}+1,\ell\right) ={}& \sqrt{2 (\ell+2)} \lambda^\frac14 -\ell-2
+ \frac{3 \ell^2+10 \ell+16}{4  \sqrt{2(\ell+2)}} \lambda^{-\frac14}\\
&-  \frac{21 \ell^4+144 \ell^3+292 \ell^2+80 \ell-128 + 96 (\ell+2)^3 \zeta(3)}{32 (2(\ell+2))^{\frac32}}
\lambda^{-\frac34} + O(\lambda^{-\frac54})\,,
\eea{tau_leading_regge}
in precise agreement with the results from integrability \cite{Gromov:2011de,Basso:2011rs,Gromov:2011bz}.

For the second Regge trajectory we find
\begin{align}
\langle f_0 \tau_2^2 \rangle_{\delta,2(\delta-2)} ={}& \frac{r_1(\delta)}{108 \delta } \left(162 \delta ^6+207 \delta ^5-376 \delta ^4+1227 \delta ^3-2156 \delta ^2+1152 \delta -648\right)\,,\nonumber\\
\langle f_0 \tau_4 + f_2 \tau_2 \rangle_{\delta,2(\delta-2)} ={}& r_1(\delta) \delta ^{5/2} \bigg( \frac19 \left(18 \delta ^3+25 \delta ^2-75 \delta +23\right) \zeta (3)
-\frac{7 \delta ^3}{6}+\frac{2941 \delta ^2}{216}-\frac{4147 \delta }{432}\nonumber\\
&-\frac{11977}{96}
+\frac{433411}{1728 \delta }
-\frac{464351}{1728 \delta ^2}
+\frac{65701}{288 \delta ^3}
-\frac{601}{8 \delta ^4}
\bigg)\,,\nonumber\\
\langle f_4 \rangle_{\delta,2(\delta-2)} ={} & r_1(\delta)  \bigg(
\frac{2}{3} \delta ^4 \left(2 \delta ^2+3 \delta -8\right) \zeta(3)^2
-\delta ^3 \left(2 \delta ^2+3 \delta -8\right)\zeta(5)\nonumber\\
&- \delta  \left(\frac{14 \delta ^5}{9}-\frac{463 \delta ^4}{27}+\frac{125 \delta ^3}{9}+\frac{41183 \delta ^2}{216}-\frac{14647 \delta }{48}+\frac{183  }{2} \right)\zeta(3)\nonumber\\
&+\frac{49 \delta ^6}{108}-\frac{31267 \delta ^5}{3240}+\frac{7109 \delta ^4}{405}+\frac{786077 \delta ^3}{12960}-\frac{3101515 \delta ^2}{5184}+\frac{49878301 \delta }{25920}\nonumber\\
&-\frac{109158059}{46080}
+\frac{97891303}{92160 \delta }
-\frac{86003}{768 \delta ^2}
\bigg)\,.
\label{OPE_data_n=1}
\end{align}
For the two operators at $\delta=2$, $\ell=0$ we can perform another consistency check. Here some of our data reads
\bea
\langle f_0 \rangle_{2,0} ={}&  \sum_{I=1}^2 f_{0}^I(2,0) = \frac14\,,\\
\langle f_0 \tau_2 \rangle_{2,0} ={}&  \sum_{I=1}^2 f_{0}^I(2,0) \tau_{2}^I(2,0) = \sqrt{2}\,,\\
\langle f_0 \tau_2^2 \rangle_{2,0} ={}&  \sum_{I=1}^2 f_{0}^I(2,0) \tau_{2}^I(2,0)^2 = 8\,,
\eea{20_data_1}
and the anomalous dimensions $\tau_{2}^I(2,0)$ were recently computed in \cite{Gromov:2023hzc}
\beq
\tau_{2}^1(2,0) = 4 \sqrt{2}\,, \qquad
\tau_{2}^2(2,0) = \sqrt{2}\,.
\eeq
We can use this input to solve the first two equations of \eqref{20_data_1} for
\beq
f_{0}^1(2,0) = \frac14\,, \qquad
f_{0}^2(2,0) = 0\,.
\eeq
The first check is that the third equation of \eqref{20_data_1} is also solved by this data.
Note that one of the operators does not enter the equations due to $f_{0}^2(2,0) = 0$. If we assume this operator to be absent at subleading orders as well, i.e.\ $f_{2}^2(2,0) = 0$, we can use
\bea
\langle f_2 \rangle_{2,0} ={}& 2 \zeta (3)-\frac{387}{256} \,,\\
\langle f_0 \tau_4 + f_2 \tau_2 \rangle_{2,0} ={}& \frac{13 \zeta (3)}{\sqrt{2}}-\frac{537}{32 \sqrt{2}} \,,
\eea{20_data_2}
to solve for
\beq
\tau_4^1(2,0) = -  \frac{75}{2 \cdot 2^{\frac32}} - \frac{24 \zeta(3)}{2^{\frac32}}  \,,
\eeq
which also agrees with the result of \cite{Gromov:2023hzc}.

\subsection{Wilson coefficients}
\label{sec:wilson_coefficients}

In order to obtain the low-energy expansion from the world-sheet integral representation
we need to compute the expansion of the function $B^{(k)}(S,T)$ defined in \eqref{B} around $S=T=0$. To this end we consider the integrals
\begin{equation}
I_w(S,T)= \int d^2z |z|^{-2S-2}|1-z|^{-2T-2} {\cal L}_w(z)\,.
\end{equation}
Their low energy expansion was computed in \cite{Alday:2023jdk}, following a method developed in \cite{Vanhove:2018elu}, with the result
\begin{equation}
I_w(S,T)= \text{polar}_w(S,T)+\sum_{p,q=0}^\infty (-S)^p (-T)^q \hspace{-15pt} \sum_{W\in 0^p \shuffle 1^q \shuffle w} \hspace{-15pt} \left( {\cal L}_{0W}(1)-{\cal L}_{1W}(1) \right).
\end{equation}
Here $\text{polar}_w(S,T)$ contains poles in $S$ and/or $T$ that originate from logarithmic divergences of the integrand and the SVMPLs at 1 can be written in terms of single-valued MZVs.
Using our solution, we find the following low-energy expansion
\begin{align}
{}& A^{(2)}(S,T) = \frac{2}{9} \frac{\hat\sigma_2^2}{\hat\sigma_3^3}
+ \frac{49}{2} \zeta(5)
+ \frac{4091}{8} \zeta(7) \hat\sigma_2 
+ 1860 \zeta(3) \zeta(5) \hat\sigma_3 
+ \frac{43664 \zeta (3)^3+343123 \zeta (9)}{108}  \hat\sigma_2^2 \nonumber\\
&+\frac{15}{16} \left(6052 \zeta (5)^2+13519 \zeta (3) \zeta (7)\right) \hat\sigma_2  \hat\sigma_3
+ \frac{309277 \zeta (11)+80144 \zeta (3)^2 \zeta (5) -108 \zeta^{\text{sv}}(5,3,3)}{24}  \hat\sigma_2^3 \nonumber\\
&+\frac{1}{128}
    \left(644933 \zeta (11)+1042048 \zeta (3)^2 \zeta (5) + 7776 \zeta^{\text{sv}}(5,3,3)\right) \hat\sigma_3^2
+ \ldots
\label{lee}
\end{align}
We provide more terms in a Mathematica notebook.
Note that \eqref{lee} includes the two Wilson coefficients 
\beq
\alpha^{(2)}_{0,0}= \frac{49}{4} \zeta(5) \,, \qquad \alpha^{(2)}_{1,0} = \frac{4091}{16} \zeta(7)\,.
\eeq
$\alpha^{(2)}_{0,0}$ was previously computed using supersymmetric localisation \cite{Chester:2020dja} and 
$\alpha^{(2)}_{1,0}$ was computed in \cite{Alday:2022xwz} by combining $A^{(1)}(S,T)$ with localisation. Both coefficients agree exactly with the previously computed values.
Having obtained the values for the Wilson coefficients $\alpha^{(2)}_{a,b}$, we can again use localisation to find all the Wilson coefficients that appear in the Mellin amplitude at order $1/\lambda^4$
\bea
{}&M(s_1,s_2) = \frac{8}{(s_1-\frac23) (s_2-\frac23) (s_3-\frac23)}
+ \frac{120 \zeta (3)}{\lambda ^{3/2}}
+ \frac{210 \left(3 \sigma _2+7\right) \zeta (5)}{\lambda ^{5/2}} \\
&+\frac{140 \left(108 \sigma_3 - 99 \sigma _2-320\right) \zeta (3)^2}{3 \lambda ^3}
+ \frac{35 \left(2592 \sigma _2^2-77328 \sigma _3+73638 \sigma _2+178909\right) \zeta (7)}{16 \lambda ^{7/2}}\\
&+\frac{10 \left(11340 \sigma _3 \sigma _2-25893 \sigma _2^2+529200 \sigma _3-473529 \sigma _2-928448\right)
   \zeta (3) \zeta (5)}{\lambda ^4}
 + O(\lambda^{-9/2})\,.
\eea{M_loc}

\section{Conclusions}

In this paper we have presented a method to compute the tree-level amplitude for four massless super-gravity states in type IIB string theory on $AdS_5 \times S^5$, order by order in the curvature corrections around flat space.  Inspired by single-valuedness and the soft theorems in superstring theory, we propose an ansatz which at each order $k$ involves (a world-sheet integration of) weight $3k$ single-valued multiple polylogarithms and a finite number of rational coefficients. The ansatz is manifestly crossing symmetric and the unknown coefficients are fixed by requiring the correct supergravity limit; the correct structure of poles, determined by dispersive sum rules; and the dimensions of the first few Konishi-like operators, available from integrability. We explicitly show how our method works for the first two curvature corrections. This can be seen as the culmination of the program started in \cite{Alday:2022uxp}, and further developed in \cite{Alday:2022xwz,Alday:2023jdk}.\footnote{See \cite{Abl:2020dbx,Aprile:2020mus} for early attempts.} There are many open problems that would be interesting to address. 
 
Our result makes very explicit the interplay between integrability and the conformal bootstrap, already elucidated in a related context in \cite{Cavaglia:2021bnz,Cavaglia:2022qpg,Caron-Huot:2022sdy}, and brings new ingredients into play, such as structures from number theory.  Some of these number theoretic structures already featured in the integrated constraints, computed via supersymmetric localisation, see {\it e.g.}\ \cite{Binder:2019jwn,Chester:2019jas,Chester:2020dja,Dorigoni:2021guq}. It would be very interesting to explore these connections further. 
 
 A framework to study open string amplitudes on $AdS$ has been introduced in \cite{Alday:2021odx} and developed in \cite{Behan:2023fqq,Glew:2023wik}. It would be interesting to study to which extent a natural map between open and closed string amplitudes in AdS exists. A related question is whether the results of \cite{Behan:2023fqq,Glew:2023wik} admit a representation in terms of a 1d integral, similar to the Veneziano amplitude, with extra insertions. 
 
Four-point tree-level amplitudes have been constructed for curved background containing $AdS$ with pure background NS-NS B-field, see \cite{Maldacena:2001km}. It would be interesting to study the low energy expansions of such amplitudes, and to understand whether single-valuedness plays a role in that case. 

The most promising formulation for a direct world-sheet approach is the pure-spinor formalism. Over the last few years there has been progress in the explicit construction of vertex operators in the pure spinor formalism \cite{Berkovits:2019rwq,Fleury:2021ieo}, but the precise integration measure to compute amplitudes is still an open problem. It would be very interesting to use the results of this paper to reconstruct said measure in a $1/R$ expansion.

\section*{Acknowledgements} 

We thank Julius Julius for useful discussions and especially Joao Silva for collaboration on related projects.
Our work is supported by the European Research Council (ERC) under the European Union's Horizon
2020 research and innovation programme (grant agreement No 787185). LFA is also supported in part by the STFC grant ST/T000864/1.

\appendix

\section{Details about dispersive sum rules}
\label{app:sum_rules}

The OPE data generally arises in the dispersive sum rules via weighted sums of the form
\beq
W_q\Big[f(\delta,\ell) \Big] = \frac{4^q}{\Gamma(2q+2)} \sum_{\ell=0,2,\ldots}^{2(\de-1)}  (\ell-q+1)_q (\ell+2)_q  f(\delta,\ell)\,.
\eeq
Concretely the OPE data enters via the functions
\begin{align}
F_q^{(1)}(\delta) ={}& W_q\Big[
 \sqrt{\delta }  \langle f_1\rangle_{\delta,\ell} - ( 3 \ell+ \tfrac{23}{4} ) \langle f_0\rangle_{\delta,\ell}   \Big]\,,\nonumber\\
T_q^{(1)}(\delta) ={}& W_q\Big[
 \langle f_0 \rangle_{\delta,\ell} (\tau_1(\delta,\ell) + \ell+2)  \Big]\,,\nonumber\\
F^{(2)}_q(\delta ) ={}& W_q\Big[    \de \langle f_2\rangle_{\delta,\ell} - \frac{39}{4} \ell \langle f_0\rangle_{\delta,\ell} \Big]\,,\nonumber\\
T^{(2)}_q(\delta ) ={}& W_q\Big[
\sqrt{\delta}\langle f_0 \tau_2 \rangle_{\delta,\ell} \Big]\,,\nonumber\\
F^{(3)}_q(\delta ) ={}& W_q\bigg[
\delta^{\frac32} \langle f_3 \rangle_{\delta,\ell}
- \delta   (3\ell+\tfrac{23}{4})
\left(\langle f_2 \rangle_{\delta,\ell} - \frac{\langle f_0 \tau_2 \rangle_{\delta,\ell}}{2 \sqrt{\delta}} \right)
+ \frac{ \ell \left(140 \ell^2+280 \ell+491\right)}{16} \langle f_0 \rangle_{\delta,\ell}
\bigg]\,,\nonumber\\
T^{(3)}_q(\delta ) ={}& W_q\Big[
\delta \langle f_0 \tau_3 \rangle_{\delta,\ell} \Big]\,,\nonumber\\
F^{(4)}_q(\delta) ={}& W_q\bigg[
\delta ^2 \<f_4\>_{\delta,\ell}
-\delta^{3/2} (3\ell+\tfrac{23}{4}) \left(\langle f_3 \rangle_{\delta,\ell} - \frac{\langle f_0 \tau_3 \rangle_{\delta,\ell}}{2 \sqrt{\delta}} \right)\nonumber\\
& \qquad +\ell \left(\frac{27}{4} \delta  \<f_2\>_{\delta,\ell}
+\frac{3}{2} \sqrt{\delta }  \<f_0 \tau_2\>_{\delta,\ell}
+ \frac{21}{32}  \left(20 \ell^2+40 \ell+137\right) \<f_0\>_{\delta,\ell} \right)\bigg]\,,\nonumber\\
T^{(4)}_q(\delta) ={}& W_q\bigg[
\delta^{\frac32} \< f_0 \tau_4 + f_2 \tau_2 \>_{\delta,\ell}
-\frac{39}{4} \ell \sqrt{\delta } \<f_0 \tau_2\>_{\delta,\ell} \bigg]\,,\nonumber\\
T^{(2,2)}_q(\delta) ={}& W_q\Big[ \delta \< f_0 \tau_2^2 \>_{\delta,\ell}\Big]\,.
\label{FT_def}
\end{align}
The dispersive sum rules further depend on the polynomials
\begin{align}
P^{(1)}_{3,0} (a,b,q) ={}&
-\frac{1}{6} (2 a+3 b)^3
+\frac{1}{6} (3 q-8) (2 a+3 b)^2
\frac{1}{12} \left(-3 q^2+52 q-2\right) (2 a+3 b)\nonumber\\
&+\frac{1}{32} \left(-216 q^2-84 q-277\right)\,,\nonumber\\
P^{(1)}_{3,1} (a,b,q) ={}& 
\frac{1}{4} (2 a+3 b)^2
+\frac{1}{24} (49-6 q) (2 a+3 b)
-\frac{3}{16} (36 q+25)\,,\nonumber\\
P^{(3/2)}_{2,0} (q) ={}&
\frac{1}{64} \left(2100 q^2+4200 q+3931\right)
\,,\nonumber\\
P^{(3/2)}_{2,1} (q) ={}&
\frac{525}{32} (2 q+3)
\,,\nonumber\\
P^{(2)}_{3,0} (a,b,q) ={}&
P^{(1)}_{3,0} (a,b,q) + \frac{1}{16} \left(144 q^2+288 q+529\right)
\,,\nonumber\\
P^{(2)}_{3,1} (a,b,q) ={}&
P^{(1)}_{3,1} (a,b,q) + \frac{9}{2} (2 q+3)
\,,\nonumber\\
P^{(2)}_{4,0} (a,b,q) ={}&
\frac{1}{6}  (2 a+3 b)^4
+\frac{1}{2} (4-q) (2 a+3 b)^3
+\frac{1}{12} \left(3 q^2-76 q+66\right) (2 a+3 b)^2\nonumber\\
&+\frac{1}{96} \left(744 q^2-1412 q+895\right) (2 a+3 b)
+\frac{3}{32} \left(240 q^2+16 q+193\right)
\,,\nonumber\\
P^{(2)}_{4,1} (a,b,q) ={}&
-\frac{1}{4} (2 a+3 b)^3
+\frac{6 q-73}{24}  (2 a+3 b)^2
+\frac{372 q-167}{48} (2 a+3 b)
+\frac{3}{2} (15 q+8)
\,,\nonumber\\
P^{(2)}_{6,0} (a,b,q) ={}&
\frac{1}{72} (2 a+3 b)^6
+\frac{1}{360} (101-30 q) (2 a+3 b)^5
+\frac{1}{36} \left(6 q^2-59 q+69\right) (2 a+3 b)^4\nonumber\\
&+\frac{1}{576} \left(-72 q^3+2304 q^2-5284 q+4455\right) (2 a+3 b)^3\nonumber\\
&+\frac{1}{576} \left(18 q^4-2688 q^3+12692 q^2-11405 q+15808\right) (2 a+3 b)^2\nonumber\\
&+\frac{1 }{5760}(9900 q^4-177180 q^3+66545 q^2-378180 q+134614)(2 a+3 b)\nonumber\\
&+\frac{1}{6144}(16128 q^4-425088 q^3-1033408 q^2-1958280 q-1285115)
\,,\nonumber\\
P^{(2)}_{6,1} (a,b,q) ={}&
-\frac{1}{24} (2 a+3 b)^5
+\frac{12 q-53}{72}  (2 a+3 b)^4
+\frac{1}{288} (-54 q^2+1098 q-763)  (2 a+3 b)^3\nonumber\\
&+\frac{1}{1152} (72 q^3-7956 q^2+17392 q-1383 )(2 a+3 b)^2\nonumber\\
&+\frac{1 }{2304}\left(7920 q^3-94428 q^2-71770 q-95783\right)(2 a+3 b)\label{P_def}\\
&+\frac{1}{1536}(8064 q^3-147312 q^2-409696 q-425081)
\,,\nonumber\\
P^{(2)}_{6,2} (a,b,q) ={}&
+\frac{1}{32} (2 a+3 b)^4
+\frac{55-6 q}{96}  (2 a+3 b)^3
+\frac{36 q^2-2616 q+613 }{1152}(2 a+3 b)^2\nonumber\\
&+\frac{1}{576}  \left(990 q^2-6879 q-8848\right) (2 a+3 b)
+\frac{1}{1536}(4032 q^2-45072 q-79693)
\,.\nonumber
\end{align}
Finally, we explain in section \ref{sec:dispersive_sum_rules} how the dispersive sum rules can be used to compute the residues of $A^{(2)}(S,T)$ at $S=\delta$ in terms of OPE data. The first four residues depend only on OPE data that has been computed previously. They are given by
\begin{align}
R^{(2)}_7(T,\delta) ={}& - \sum\limits_{q=0}^{2(\delta-1)} \left( \frac{T}{\delta}\right)^q 10 \delta^2 F^{(0)}_{q}(\delta) \,, \nonumber\\
R^{(2)}_6(T,\delta) ={}& - \sum\limits_{q=0}^{2(\delta-1)} \left( \frac{T}{\delta}\right)^q \delta
\left(\frac{4}{3}   F^{(0)}_{q}(\delta)+5   (q+1) F^{(0)}_{q+1}(\delta) \right)\,,\nonumber\\
R^{(2)}_5(T,\delta) ={}& \sum\limits_{q=0}^{2(\delta-1)} \left( \frac{T}{\delta}\right)^q \bigg(
\frac{1}{3} \left(3 q^2+17 q+25\right) F^{(0)}_{q}(\delta)
+\frac{1}{3} (q+1) (3 q+8) F^{(0)}_{q+1}(\delta)\nonumber\\
&-\frac{3}{4} (q+1) (q+2) F^{(0)}_{q+2}(\delta)
-4 T^{(2)}_{q}(\delta)
\bigg)\,,\nonumber\\
R^{(2)}_4(T,\delta) ={}& \sum\limits_{q=0}^{2(\delta-1)} \left( \frac{T}{\delta}\right)^q \frac{1}{\delta} \bigg(
\frac{1}{96} \left(-8 q^3+384 q^2+492 q+1119\right) F^{(0)}_{q}(\delta)\nonumber\\
&+\frac{1}{48} (q+1) \left(18 q^2+370 q+577\right) F^{(0)}_{q+1}(\delta)
+\frac{1}{16} (q+1)_2 (6 q+25) F^{(0)}_{q+2}(\delta)\nonumber\\
&+(q+2) T^{(2)}_{q}(\delta)
-\frac{3}{2} (q+1) T^{(2)}_{q+1}(\delta)
-F^{(2)}_{q}(\delta)
\bigg)\,.
\label{R_from_dsr_7-4}
\end{align}
The remaining residues depend on the new OPE data in $T^{(2,2)}_q(\delta)$, $T_q^{(4)}(\delta)$ and $F_q^{(4)}(\delta)$ and are given by
\begin{align}
{}&R^{(2)}_3(T,\delta) = \sum\limits_{q=0}^{2(\delta-1)} \left( \frac{T}{\delta}\right)^q \frac{1}{\delta^2} \bigg(
\frac{1}{144} \left(-9 q^4-112 q^3-1000 q^2-1884 q-2271\right) F^{(0)}_{q}(\delta)\nonumber\\
&-\frac{q+1}{576}  \left(96 q^3-300 q^2+1036 q+49\right) F^{(0)}_{q+1}(\delta)
-\frac{(q+1)_2}{576}  \left(36 q^2-1068 q-2447\right) F^{(0)}_{q+2}(\delta)\nonumber\\
&
+\frac{3 q^2+16 q+28}{6}  T^{(2)}_{q}(\delta)
+\frac{q+1}{12}  (12 q+37) T^{(2)}_{q+1}(\delta)
+\frac{2}{3} F^{(2)}_{q}(\delta)
-\frac{q+1}{2}  F^{(2)}_{q+1}(\delta)
-T^{(2,2)}_{q}(\delta)
\bigg)\,,\nonumber\\
{}&R^{(2)}_2(T,\delta) = \sum\limits_{q=0}^{2(\delta-1)} \left( \frac{T}{\delta}\right)^q \frac{1}{\delta^3} \bigg(
\frac{1}{1152} (q+4) (3 q+8) \left(8 q^3-384 q^2-428 q-927\right) F^{(0)}_{q}(\delta)\nonumber\\
&+\frac{1}{2304} (q+1) \left(48 q^4-4912 q^3-42956 q^2-111278 q-107749\right) F^{(0)}_{q+1}(\delta)\nonumber\\
&-\frac{5 }{1152}(q+1)_2 \left(276 q^2+2024 q+3091\right)  F^{(0)}_{q+2}(\delta)
+\frac{1}{96} \left(-32 q^3+176 q^2-148 q+415\right) T^{(2)}_{q}(\delta)\nonumber\\
&-\frac{1}{48} (q+1) \left(12 q^2-152 q-187\right) T^{(2)}_{q+1}(\delta)
+\frac{1}{12} (q+4) (3 q+8) F^{(2)}_{q}(\delta)\nonumber\\
&+\frac{1}{24} (q+1) (6 q+31) F^{(2)}_{q+1}(\delta)
+\frac{1}{4} (4 q+7) T^{(2,2)}_{q}(\delta)
-T^{(4)}_{q}(\delta)
\bigg)\,,\nonumber
\end{align}
\begin{align}
{}&R^{(2)}_1(T,\delta) = \sum\limits_{q=0}^{2(\delta-1)} \left( \frac{T}{\delta}\right)^q \bigg(
\frac{1}{92160} (-320 q^6+30144 q^5+1157440 q^4+5790960 q^3+18048880 q^2\nonumber\\
&+25797336 q+19462005)   F^{(0)}_{q}(\delta)
+\frac{q+1 }{4608} (1904 q^4+98248 q^3+504424 q^2+1237298 q\nonumber\\
&+1080201) F^{(0)}_{q+1}(\delta)
-\frac{(q+1)_2 }{4608} \left(96 q^3-35252 q^2-158704 q-168339\right) F^{(0)}_{q+2}(\delta)\nonumber\\
&+\frac{1}{96} \left(8 q^4-360 q^3-1148 q^2-1347 q-1194\right) T^{(2)}_{q}(\delta)
-\frac{q+1}{48}  \left(238 q^2+809 q+609\right) T^{(2)}_{q+1}(\delta)\nonumber\\
&+\frac{1}{96} \left(-8 q^3-480 q^2-1300 q-2247\right) F^{(2)}_{q}(\delta)
-\frac{1}{48} (q+1) (194 q+337) F^{(2)}_{q+1}(\delta)\nonumber\\
&-\frac{1}{4} (q+2) (2 q+5) T^{(2,2)}_{q}(\delta)
+(q+2) T^{(4)}_{q}(\delta)
-F^{(4)}_{q}(\delta)
\bigg)\,.
\label{R_from_dsr_3-1}
\end{align}

\section{Single-valued multiple polylogarithms}
\label{app:SVMPLs}
A crucial role in our construction is played by single-valued multiple polylogarithms. Let us first introduce multiple polylogarithms (MPLs), also denoted harmonic polylogarithms. These are functions of a single variable $L_w(z)$ labelled by a word $w$ in the alphabet $\{0,1\}$. They can be defined recursively by 
\begin{equation}
\frac{d}{dz} L_{0w}(z) = \frac{1}{z} L_{w}(z)\,,\qquad
\frac{d}{dz} L_{1w}(z) = \frac{1}{z-1} L_{w}(z),
\end{equation}
together with the condition $\lim_{z \to 0} L_w(z)=0$ unless $w=0^p$ for which $L_{0^p}(z) = \frac{\log^p z}{p!}$. In particular for the empty word we have $L_\emptyset(z)=1$. For instance 
\begin{equation}
L_{0^{n-1}1}=-\text{Li}_n(z)\,,
\end{equation}
reduce to the classical polylogarithms. MPLs can also be given in terms of nested integrals since
\begin{equation}
L_{0w}(z) = \int_0^z \frac{1}{z'} L_{w}(z') dz'\,, \qquad
L_{1w}(z) = \int_0^z \frac{1}{z'-1} L_{w}(z') dz'.
\label{nestedint}
\end{equation}
MPLs satisfy various relations, in particular the shuffle-relations
\begin{equation}
L_{w}(z) L_{w'}(z)  = \sum_{W \in w \shuffle w'} L_{W}(z)\,.
\label{shuffle}
\end{equation}
At $z=1$ they define multiple zeta values $L_{w}(1) = \zeta(w)$, using a non-standard notation for $\zeta(w)$ where $w$ denotes the word labelling the  multiple polylogarithm under consideration. In this definition, logarithmic divergences as $z \to 1$ are isolated by using the shuffle relations and regulated by defining $\zeta(1)=0$. MPLs are closed under the transformations
\begin{equation}
z \to 1-z,~~~~z \to \frac{z}{z-1},
\end{equation}
and compositions of those. To work out their transformation properties is tedious but in principle straightforward (see for instance \cite{Maitre:2005uu}). Let's start with $z \to 1-z$. For weight one we have
\begin{equation}
L_{0}(1-z) = L_{1}(z),~~~ L_{1}(1-z) = L_{0}(z)\,.
\end{equation}
For higher weight we can proceed recursively. If a word starts with $0$ one can easily show, using the integral representation
\begin{equation}
L_{0w}(1-z) = L_{0w}(1) + \int_0^z \frac{1}{z'-1} L_{w}(1-z') dz'\,,
\end{equation}
with $L_{w}(1-z')$ given in terms of $L_{w'}(z')$ by the transformations at lower order. If a word starts with $1$ we can always use the shuffle identities to express $L_{1w}(z)$ in terms of $L_{1}(z)$ (whose transformation properties are known) and MPLs whose words start with 0. The final transformation properties take the form
\begin{equation}
L_{w}(1-z) = L_{f \cdot w}(z) + \zeta's \times\text{lower weight MPLs}\,,
\end{equation}
where $f \cdot w$ denotes the word $w$ after flipping $0 \leftrightarrow 1$ in each place. Let us now focus on the second transformation. At weight one we have 
\begin{equation}
L_{0}\left(\frac{z}{z-1} \right) = L_{0}(z)-L_{1}(z) \pm i \pi\,,~~~ L_{1}\left(\frac{z}{z-1} \right)= -L_{1}(z) \,.
\end{equation}
From now on we will disregard the $\pm i \pi$ as ultimately we are interested in the transformation properties of the single-valued versions of MPLs, to be defined below. At higher weight we can proceed recursively and obtain
\bea
L_{0w}\left(\frac{z}{z-1} \right) &= \int_0^z \left(\frac{1}{z'}-\frac{1}{z'-1} \right) L_w \left(\frac{z'}{z'-1} \right)\,, \\
L_{1w}\left(\frac{z}{z-1} \right) &=- \int_0^z \frac{1}{z'-1} L_w \left(\frac{z'}{z'-1} \right)\,.
\eea{Lint}
Plugging the transformation properties of $L_w \left(\frac{z'}{z'-1} \right)$, known by assumption, we obtain those of higher weight. 

We will also be interested in the expansions of MPLs around $z=0$. Non analytic terms, that go like $\log^n z$, arise whenever the word $w$ ends in one or more zeros. These logarithmic terms can be isolated by using the shuffle relations. For instance
\begin{equation}
L_{10}(z) = L_1(z) L_0(z) - L_{01}(z)\,,
\end{equation}
so that we can focus on words ending in 1. In this case MPLs are analytic around $z=0$ and we have
\begin{equation}
L_{w}(z) = \sum_{\ell=1}^\infty c_w(\ell) z^\ell\,.
\end{equation}
Plugging this into the integral representations leads to the following recursive relations
\begin{equation}
c_{0w}(\ell) = \frac{c_{w}(\ell)}{\ell},~~~ c_{1w}(\ell) = -\sum_{\ell'=1}^{\ell-1} \frac{c_w(\ell')}{\ell}\,,
\end{equation}
which together with $c_1(\ell)=-1/\ell$ fix all $c_w(\ell)$ recursively. These recursions can be solved in terms of Euler-Zagier sums.

In the complex $z-$plane multiple polylogarithms are analytic functions with branch points at $z=0,1$ ( and the point at infinity if we consider the Riemann sphere). It is possible to construct single-valued multiple polylogarithms (SVMPLs) ${\cal L}_w(z)$ which are weight preserving linear combinations of $L_{w_1}(z)L_{w_2}(\bar z)$ such that all discontinuities cancel. They are defined such that they satisfy the same differential relations 
\begin{equation}
\frac{\partial }{ \partial z} {\cal L}_{0w}(z) = \frac{1}{z} {\cal L}_{w}(z),~~~~~\frac{\partial}{\partial z} {\cal L}_{1w}(z) = \frac{1}{z-1} {\cal L}_{w}(z),
\end{equation}
together with the condition $\lim_{z \to 0} {\cal L}_w(z)=0$ unless $w=0^p$ for which ${\cal L}_{0^p}(z) = \frac{\log^p z \bar z}{p!}$. Furthermore, SVMPLs satisfy the same shuffle relations 
\begin{equation}
{\cal L}_{w}(z) {\cal L}_{w'}(z)  = \sum_{W \in w \shuffle w'} {\cal L}_{W}(z).
\label{shufflesv}
\end{equation}
and at $z=\bar z=1$ they define single-valued multiple zeta values. Their explicit construction in terms of MPLs was presented for instance in \cite{Brown:2004ugm,Dixon:2012yy}. In particular, from these explicit constructions, plus the relations worked out above, one can find their transformation properties under $z \to 1-z$ and $z \to \frac{z}{z-1}$ as well as their expansions around $z,\bar z=0$.

\bibliographystyle{JHEP}
\bibliography{A2}
\end{document}